\documentclass[12pt]{article}
\usepackage{amsmath,amssymb,graphicx} 
\usepackage{epsf,color,hyperref}
\usepackage{cite}

\newcommand{\beq}{\begin{eqnarray}}
\newcommand{\eeq}{\end{eqnarray}}
\newcommand{\centeron}[2]{{\setbox0=\hbox{#1}\setbox1=\hbox{#2}\ifdim

\wd1>\wd0\kern.5\wd1\kern-.5\wd0\fi
\copy0

\kern-.5\wd0\kern-.5\wd1\copy1\ifdim\wd0>\wd1
                                       \kern.5\wd0\kern-.5\wd1\fi}}
\newcommand{\ltap}{\>\centeron{\raise.35ex\hbox{$<$}}
                               {\lower.65ex\hbox{$\sim$}}\>}
\newcommand{\gtap}{\>\centeron{\raise.35ex\hbox{$>$}}
                               {\lower.65ex\hbox{$\sim$}}\>}

\newcommand\ZZ{\hbox{\zfont Z\kern-.4emZ}}
\font\zfont = cmss10 

\begin{document}
\begin{titlepage}
\begin{flushright}
\end{flushright}

\vskip.5cm
\begin{center}
{\huge \bf Anomaly Constraints on Monopoles and Dyons\\}

\vskip.1cm
\end{center}
\vskip0.2cm

\begin{center}
{\bf Csaba Cs\'aki$^{a}$,
Yuri Shirman$^{b}$, {\rm and}
John Terning$^{c,d}$}
\end{center}
\vskip 8pt

\begin{center}
$^{a}$ {\it Institute for High Energy Phenomenology\\
Newman Laboratory of Elementary Particle Physics\\
Cornell University, Ithaca, NY 14853, USA } \\
$^{b}$ {\it
Department of Physics, University of California, Irvine, CA
92617.} \\

$^{c}$ {\it
Department of Physics, University of California, Davis, CA
95616.} \\

$^{d}$ {\it
CERN, Physics Department, Theory Unit, Geneva, Switzerland.}\\

\vspace*{0.3cm}
{\tt   csaki@cornell.edu, shirman@uci.edu, terning@physics.ucdavis.edu}
\end{center}

\vglue 0.3truecm

\begin{abstract}
\vskip 3pt \noindent Fermions with magnetic charges can contribute to anomalies. We derive the axial anomaly and gauge anomalies for monopoles and dyons, and find eight new gauge anomaly cancelation conditions in a general theory with both electric and magnetic charges.  As a byproduct we also extend the Zwanziger two-potential formalism to include the $\theta$ parameter, and elaborate on the condition for CP invariance in theories with fermionic dyons. 
\end{abstract}

\end{titlepage}

\newpage


\section{Introduction}
\label{sec:intro}
\setcounter{equation}{0}
\setcounter{footnote}{0}

It is well known that there is a
charge quantization constraint in a $U(1)$ gauge theory with both electric and magnetic charges \cite{Dirac,Zwanziger:1968rs,Schwinger,MonopoleReviews}.  If we label the electric and magnetic charges of particle $j$ by $q_j$  (measured in units of the coupling $e$) and $g_j$  (measured in units of $4 \pi/e$)  then the charges of any pair of particles must satisfy
\beq
q_i g_j-q_j g_i=\frac{ n}{2}~,
\eeq
where $n$ is an integer that can be different for each pair.
In a CP invariant theory this requires that both types of charges can be expressed as integers in units of a fundamental  charge \cite{Witten:1979ey}.
There are also five well known conditions on electric $U(1)$ gauge charges of fermions that arise from requiring anomaly cancelations.
The standard gauge anomaly conditions come from the $U(1)^3$ gauge anomaly, as well as the various mixed anomalies between the $U(1)$ and other possible force carriers. In general, these are  the $SU(N)^2 U(1)$ mixed anomaly,  the $U(1)_X U(1)^2$ mixed anomaly, the $U(1) U(1)_X^2$ mixed anomaly, and the  mixed gravitational $U(1)$ anomaly.  We can write these conditions, in order, as:
\beq
\sum_j  q_j^3=0~,\label{cubicanomaly}\\
\sum_j {\rm Tr}\,T^a_{r_j} T^b_{r_j} q_j\equiv \delta^{ab}\,\sum_j T(r_j) q_j=0~,\\\sum_j q_{Xj} q_j^2=0~,\label{SUN^2U1anomaly}\\
\sum_j q_{Xj}^2 q_j=0~,\label{U1X^2U1anomaly}\\
\sum_j q_j=0\label{gravanomaly}~.
\eeq
From the work of Seiberg and Witten \cite{SeibergWitten} we know that there are consistent theories with massless fermionic magnetic monopoles.
We are thus led to ask a very simple question: are there not also anomaly conditions on magnetic charges?  

We expect that a consistent theory with magnetically charged  topological solitons will give a consistent low-energy theory, but here we are asking a bottom-up question: what are the possible consistent low-energy effective field theories involving massless fermionic monopoles.  The monopoles may be fundamental or they may be topological solitons, as long as they are light compared to the inverse of their physical size we would hope to be able to write an effective theory for them.

The first hint that there may indeed be non-trivial anomaly constraints arises in a theory with a dyon, i.e. a particle with both $q_j$ and $g_j$ non-vanishing \cite{Schwinger:1969ib},  and a CP violating $\theta$ parameter. As was shown by  Witten \cite{Witten:1979ey} the effective electric charge (in units of the fundamental charge) becomes
\beq
q_{{\rm eff},j} = q_j+g_j \frac{\theta}{2 \pi}
\eeq
Disregarding cancellations that occur for particular values of $\theta$ one might naively expect that we get  new anomaly cancellation conditions for the magnetic charges by replacing $q_j$ by $q_{{\rm eff},j}$ in Eqs.~(\ref{cubicanomaly})-(\ref{gravanomaly}) and requiring that terms with different powers of $\theta$ vanish independently.

This argument is too naive, for two reasons.  First, since the magnetic charge also couples to the electromagnetic field,  there should be additional contributions proportional to powers of the magnetic charge even with $\theta=0$.  Secondly as the mass of a charged fermion goes to zero, the $\theta$ dependent piece of the charge becomes delocalized \cite{Callan:1982au}, and $\theta$ becomes an unphysical parameter at zero mass.
In what follows we will find both the electric and magnetic contributions and find new conditions that must be satisfied even in a CP conserving theory with $\theta=0$.

The paper is organized as follows. In Section 2 we give a brief review of $SL(2,Z)$ dualities and how they can be employed to easily calculate $\beta$-functions for dyons. In Section 3 we extend Zwanziger's two-potential formalism to incorporate a non-vanishing $\theta$-parameter. This gives a local, but non-Lorentz invariant, Lagrangian description of dyons where the $SL(2,Z)$ duality is explicit. In Section 4 we discuss the issue of CP invariance in theories with fermionic dyons. In Section 5 we calculate the axial anomaly from dyons, and finally in Section 6 we present the complete set of anomaly cancelation conditions for a $U(1)$ gauge theory with dyons and other possible gauge interactions.

\section{Review of $SL(2,Z)$ and $\beta$-functions}
\label{sec:SL2Z}
\setcounter{equation}{0}
\setcounter{footnote}{0}

Since it is impossible to write a local and Lorentz invariant Lagrangian for coexisting monopoles and dyons, direct loop calculations are quite difficult to perform. One of the main tools we will be using here to circumvent this problem are $SL(2,Z)$ duality transformations. Thus it is important to be very clear what the exact meaning of these transformations is. There exist some very special theories (usually ${\cal N}=2$ or ${\cal N}=4$ superconformal theories) which have a manifest$SL(2,Z)$ symmetry, which means that the entire particle spectrum is invariant under $SL(2,Z)$. Here we will {\em not} be confining ourselves to such theories, and we will be using $SL(2,Z)$ in a different way, merely as a set of field redefinitions. For us $SL(2,Z)$ will be just a particular change of variables. Let us review in detail how this comes about~\cite{ArgyresSUSY}. Consider a U(1) gauge theory with coupling $e$ and a non-vanishing $\theta$ angle in the non-canonical (``holomorphic") normalization of the gauge fields:
\beq
{\mathcal L}_{\rm free}=-\frac{1}{4 e^2} F^{\mu \nu}F_{\mu \nu} -\frac{\theta}{32 \pi^2} F^{\mu \nu}\,^*F_{\mu \nu}
\eeq
where
\beq
^*F^{\mu \nu}=\frac{1}{2}\epsilon^{\mu\nu\alpha\beta} F_{\alpha \beta} ~.
\label{eq:dualF}
\eeq
It is very convenient to introduce the holomorphic gauge coupling $\tau$, defined as
\beq
\tau \equiv {{\theta}\over{2 \pi}} +{{4 \pi i}\over{e^2}}~.
\eeq
With this notation the Lagrangian of the free theory (without any electric or magnetic charges) can be rewritten as:
\beq
{\mathcal L}_{\rm free}=-{\rm Im}\, \frac{ \tau}{32 \pi}  \left( F^{\mu \nu}+ i \,^*F^{\mu \nu}\right)^2~.
\eeq
One can see, that a shift in $\tau$ by a real integer $\tau \to \tau + n$ corresponds to shifts in the $\theta$ angle $\theta \to \theta + 2\pi n$. This is often referred to as a T-duality. Even though this does not leave the Lagrangian invariant, it is a symmetry of the theory since the only way the path integral depends on $\theta$ is via the phase $e^{i m \theta}$. To obtain the full $SL(2,Z)$ transformation group one also needs to introduce the duality field transformations. This is nothing but a change of variables in the path integral (for example nicely described in~\cite{ArgyresSUSY}).  The main point is that the path integral in terms of the (electric) gauge potential $A_\mu$ given by
\begin{equation}
\int DA_\mu e^{iS}
\end{equation}
can be thought of as a path integral in terms of the field strength $F_{\mu\nu}$ if a spin 1 Lagrange multiplier $B_\mu$ is added to enforce the Bianchi identity
\begin{equation}
\mathcal{L}_c=\frac{1}{4\pi} \int d^4 B_\mu \partial_\nu\,^*F^{\mu\nu}.
\end{equation}
The field $F$ can be integrated out from the action ${\mathcal L}_{\rm free}+{\mathcal L}_c$, and the resulting action is given by
\beq
\tilde{\mathcal{L}}={\rm Im}\, \frac{1}{32 \pi \tau}  \left( \tilde{F}^{\mu \nu}+ i \,^*\tilde{F}^{\mu \nu}\right)^2~,
\eeq
where $\tilde{F}_{\mu\nu}=\partial_\mu B_\nu-\partial_\nu B_\mu$. Thus there is a  duality with $\tau \to -\frac{1}{\tau}$, which is usually referred to as S-duality. Note, that  S-duality also has the effect of exchanging electric and magnetic charges with each other. This follows from the fact that the magnetic charge would show up as a source in the Bianchi identity, and thus would couple to the Lagrange multiplier field $B_\mu$, which becomes the electric field in the dual description. Combining the T and S-dualities we obtain the full $SL(2,Z)$ group under which the coupling transforms as
\beq
\tau^\prime =\frac{a \tau +b}{c \tau +d}~.
\label{eq:tautrans}
\eeq
with $a,b,c,d$ integers satisfying $ad-bc=1$.
Under an $SL(2,Z)$ transformation, the magnetic current $K^\mu$ and the electric current $J^\mu$
are mapped to
\beq
K^\mu \rightarrow a K^{\prime \mu} + c J^{\prime \mu}~,\,\,\,   J^\mu \rightarrow b K^{\prime \mu} + d J^{\prime \mu}~.
\label{sl2zcharge}
\eeq
This can be seen by requiring that under T the Witten charge $q+\frac{\theta}{2\pi} g$ remains invariant, together with the exchange electric and magnetic fields  under S-duality.

Let us look at a simple application  of these $SL(2,Z)$ transformations by calculating the $\beta$-function of a theory with arbitrary monopoles and dyons. This was first presented by Argyres and Douglas in~\cite{ArgyresDouglas}. The perturbative $\beta$-function is defined by
\beq
\frac{d\tau}{d \log \mu} = \beta .
\eeq
Assume that we have a dyon with electric charge $q$ and magnetic charge $g$ (in units of $e$ and $4\pi/e$). To find the $\beta$-function, we do an $SL(2,Z)$ transformation to a basis where the dyon has a pure electric charge $n$, where $n$ is the greatest common divisor of $q$ and $g$, i.e. $n=$gcd$(q,g)$. This can be achieved via the $SL(2,Z)$ transformation of the form
\beq \left( \begin{array}{cc} a & -b \\ -c & d \end{array} \right) \left( \begin{array}{c} q \\ g \end{array}\right)=
\left( \begin{array}{c} n \\ 0 \end{array}\right),
\label{eq:dyontrans}
\eeq
with the choice $c=g/n, d=q/n$ and $a,b$ integers that satisfy $a q-bg=n$. In this new transformed basis the coupling is given by (\ref{eq:tautrans}). The one loop $\beta$-function is well-known to be
\beq
\frac{d\tau'}{d\log \mu} = i \frac{n^2}{16\pi^2}.
\eeq
Rewriting this in terms of $\tau$ and using the explicit expressions for $a,b,c,d$ we find
\beq
\frac{d\tau}{d\log \mu} = \frac{i}{16\pi^2}  (q+g\tau)^2 .
\eeq
Separating out the real and imaginary parts we can obtain the separate $\beta$-functions for the gauge coupling and the $\theta$ angle (assuming the presence of several dyons at the same time):
\beq
\label{betae}
\beta_{e}&=& \mu {{d e}\over{d \mu}} =  \frac{e^3}{12\pi^2}\sum_j \left[ \left(q_j+\frac{\theta}{2 \pi} g_j\right)^2-g_j^2 \frac{16 \pi^2}{e^4}\right]~,\\
\beta_{\theta}&=& \mu {{d \theta}\over{d \mu}} =  -\frac{16 \pi }{3}\sum_j  \left[g_j \left(q_j + \frac{\theta}{2 \pi} g_j\right)\right]~.
\label{betatheta}
\eeq
This agrees with the one-loop $\beta$-functions calculated using perturbative methods
\cite{Laperashvili:1999pu}.

\section{The Generalized Zwanziger Lagrangian}
\label{sec:zwans}
\setcounter{equation}{0}
\setcounter{footnote}{0}

To perform our anomaly calculations we will need 
to determine $SL(2,Z)$ transformations properties of the gauge field.  The transformation of the field strength can be extracted relatively easily
from
the requirement that the equations of motion are covariant under $SL(2,Z)$. To write down the correct Maxwell equations we need to also incorporate the Witten effect: for non-vanishing $\theta$ the magnetic current also couples to the electric fields. An effective Maxwell equation correctly reproducing this is given by (using our normalizations):
\beq
\frac{{\rm Im}\left(\tau\right)}{4 \pi} \, \partial_\mu \left( F^{\mu \nu}+ i \,^*F^{\mu \nu}\right) = J^\nu+ \tau K^\nu~.
\label{eq:maxwell}
\eeq
The transformations of the currents in (\ref{sl2zcharge}) can be combined with the mapping
\beq
\left( F^{\mu \nu}+ i \,^*F^{\mu \nu}\right)
\rightarrow \frac{1}{c \tau^* +d}\left( F^{\prime \mu \nu}+ i \,^*F^{\prime \mu \nu}\right)
\label{fsl2z}
\eeq
to find that the effective Maxwell equation (\ref{eq:maxwell}) are covariant under the $SL(2,Z)$ duality transformations: the dual equations of motion have exactly the same form in terms of the dual coupling as the original equations:
\beq
\frac{{\rm Im}\left( \tau^\prime \right)}{4 \pi} \,  \partial_\nu \left( F^{\prime \mu \nu}+ i \,^*F^{\prime \mu \nu}\right) = J^{\prime \mu}+ \tau^\prime K^{\prime \mu}~.
\eeq

However we will also need to know how the gauge potentials transform under $SL(2,Z)$, which is a little more subtle, since it requires knowledge of the action.
It is well known that it is impossible to write a local, Lorentz invariant Lagrangian for a $U(1)$ theory with both electric and magnetic charges. Dirac originally wrote down a non-local, Lorentz invariant Lagrangian \cite{Dirac:1948um} and later Zwanziger \cite{Zwanziger:1970hk} was able to reformulate the theory in terms of a local, non-Lorentz invariant Lagrangian with two gauge potentials $A_\mu$ and $B_\mu$. Even though there are two gauge potentials, the form of the non-Lorentz invariant kinetic mixing ensures that the are only two on-shell degrees of freedom for the gauge fields. The advantage of having two gauge potentials is that one, $A_\mu$, has a local coupling to electric currents, while $B_\mu$ has a local coupling to magnetic currents.  In Dirac's formulation, the magnetic current does not couple directly to the gauge field, it only couples through the Dirac string attached to each monopole, which makes calculations very difficult.

For our work we will need to generalize the Zwanziger action to include the CP violating parameter
$\theta$. The use of differential forms also makes the expressions slightly easier to write, so we will use the notation
\beq
(a\wedge b)^{\mu\nu}= a^\mu b^\nu-b^\mu a^\nu~,\\
\left(a \cdot \,^* (b\wedge c)\right)^\nu= \epsilon^{\mu \nu \alpha \beta} a_\mu b_\alpha c_\beta~.
\eeq
Zwanziger found~\cite{Zwanziger:1970hk} that for $\theta=0$ the Maxwell equations are reproduced by the action 
\beq
\label{eq:Zwanziger}
{\mathcal L}&=& -\frac{1}{2 n^2 e^2} \left\{ \left[ n\cdot (\partial \wedge A)\right] \cdot \left[ n\cdot ^*(\partial \wedge B)\right]-\left[ n\cdot (\partial \wedge B)\right] \cdot \left[ n\cdot ^*(\partial \wedge A)\right]\right. \nonumber \\ && \left.+\left[ n\cdot (\partial \wedge A)\right]^2+\left[ n\cdot (\partial \wedge B)\right]^2\right\} - J\cdot A -\frac{4\pi}{e^2} K\cdot B \,,
\eeq
where $n$ is an arbitrary four vector corresponding to the direction of the Dirac string and
the field strength $F$ is given by
\beq
F=\frac{1}{n^2}\left(\left\{n \wedge \left[n\cdot (\partial \wedge A)\right]\right\}-\, ^*\left\{n \wedge \left[n\cdot (\partial \wedge B)\right]\right\}\right)~.
\label{Feqm}
\eeq
While the Lagrangian is not Lorentz invariant, the EOM's are Lorentz covariant if written in terms of the field strength, as in Eq. (\ref{eq:maxwell}). 

The proper generalization of this Lagrangian incorporating the $\theta$-angle is
\beq
{\mathcal L}&=&-{\rm Im} \frac{ \tau}{8\pi n^2} \left\{ \left[ n\cdot\partial \wedge (A+iB)\right]\cdot \left[ n\cdot \partial \wedge (A-iB)\right]\right\} \nonumber \\ &&-{\rm Re} \frac{ \tau}{8\pi n^2} \left\{ \left[ n\cdot\partial \wedge (A+iB)\right]\cdot \left[ n\cdot ^*\partial \wedge (A-iB)\right]\right\} \nonumber \\ &&- J\cdot A -\frac{4\pi}{e^2} K\cdot B .
\label{eq:fullZwanziger}
\eeq
One can check that this Lagrangian indeed correctly reproduces the Maxwell equations (\ref{eq:maxwell}) after the Witten effect is taken into account. To incorporate the Witten effect, one may also write a low-energy Lagrangian below the mass scale of the fermions that will correct the coupling terms to
\begin{equation}
\label{eq:effcoupling}
- J\cdot A -\frac{4\pi}{e^2} K\cdot B  \to {\rm Re} \left[ (A-i B) \cdot (J+\tau K)\right] 
\end{equation}
while in the case of massless fermions the $\theta$ term can always be rotated away. 

One can easily see that, with this incorporation of the Witten effect in the coupling of the Lagrangian, the $SL(2,Z)$ covariance is also explicit. Since under $SL(2,Z)$ the field strength should transform as $\left( F+ i \,^*F\right)
\rightarrow \frac{1}{c \tau^* +d}\left( F^{\prime}+ i \,^*F^{\prime}\right)$, with the identification (\ref{Feqm}) one expects that the proper $SL(2,Z)$ transformation of $A,B$ is
\beq
\left(A+i B\right)\rightarrow \frac{1}{c \tau^* +d}\left(A^\prime+i B^\prime\right)~.
\label{gaugepotSL2Z}
\eeq
One can check that the Lagrangian (\ref{eq:fullZwanziger}) with the modification in (\ref{eq:effcoupling}) is indeed covariant under the combined transformation (\ref{sl2zcharge}) and (\ref{gaugepotSL2Z}).

\section{CP}
\label{sec:cp}
\setcounter{equation}{0}
\setcounter{footnote}{0}

If the $\theta$ parameter vanishes we can have a CP invariant theory, provided that the spectrum
of monopoles and dyons is CP invariant.  The details of the particle spectrum is not something we have to consider when checking CP invariance for ordinary charged particles (scalars or fermions). For example for a 2-component Weyl spinor $e_\alpha$ CP takes a left-handed electron field to left-handed positron:
\beq
 e_\alpha \rightarrow  \sigma^{2}_{\alpha{\dot\alpha}}  e^{\dagger {\dot \alpha}}~.
 \eeq
Since the CP conjugate is just the hermitian conjugate,
any set of electrically charged fields has a CP invariant spectrum.  However, the electric field $\vec{E}$ and the magnetic field $\vec{B}$ have opposite CP, so  the CP conjugate of a particle with a magnetic charge $g$ also must have magnetic charge $g$, not $-g$. This fact has often been quoted in the literature as a reason for theories with dyons to necessarily break CP~\cite{Jackson}. However Witten emphasized that this does not have to be the case, if one can modify the definition of CP such that it also includes the exchange of different fields. However, this is only possible if a particular pairing among the charges of the fields holds, which will lead to  restrictions for the possible charges for dyons.  Suppose that we have a Weyl fermion $\chi$ with electric and magnetic charges given by $(q, g)$. For the theory to be CP invariant we need another Weyl dyon $\psi$ with charges $(q, -g)$ so that  CP can interchange the fields $\chi$ and $\psi^\dagger$ rather than replacing them by their own hermitian conjugates.

This can be nicely incorporated into the Zwanziger's two potential formalism used in the previous section. In this formalism the electric charge couples to the $A$ field and the magnetic charge to $B$, so we can simply write the gauge couplings of the two dyons mentioned above as
 \beq
 {\mathcal L}_{\rm int}= -\chi^\dagger \left(q \,A_\mu + \tilde g \, B_\mu\right){\bar \sigma}^\mu \chi
  -\psi^\dagger \left(q \,A_\mu - \tilde g\, B_\mu\right){\bar \sigma}^\mu \psi~,
  \label{dyonint}
  \eeq
where we have defined $\tilde g\equiv g \,4 \pi/e^2$.
 Now $A_\mu$ and $B_\mu$ must have opposite CP, this can be seen from the fact that one couples to electric charge and one couples to magnetic charge, or from the fact that their kinetic mixing (in the absence of the $\theta$-term, see (\ref{eq:Zwanziger}) involves the  pseudo-tensor $\epsilon^{\mu\nu\alpha\beta}$ which is odd under CP.
 Thus the CP transformation of the gauge fields are:
 \beq
 A_0 \rightarrow - A_0~,\,\,\, A_i \rightarrow +A_i~,\,\,\,  B_0 \rightarrow + B_0~,\,\,\, B_i \rightarrow -B_i~,
 \eeq
while the dyons transform under CP as
\beq
\chi_\alpha \rightarrow \sigma^{2}_{\alpha{\dot\alpha}} \psi^{\dagger {\dot \alpha}}~,\,\,\,\psi_\alpha \rightarrow \sigma^{2}_{\alpha{\dot\alpha}} \chi^{\dagger {\dot \alpha}}\, .
\eeq
Since fermions anticommute, we have:
\beq
&&\chi^\dagger \left(q \,A_0 + \tilde g \, B_0 \right) \chi\rightarrow \psi \sigma^2 \left(q \,(-A_0)+ \tilde g \, B_0 \right) \sigma^2 \psi^\dagger  =
\psi^\dagger \left(q \,A_0- \tilde g \, B_0 \right) \psi \\
&&\chi^\dagger \left(q \,A_j+ \tilde g \, B_j \right){\bar \sigma}^j \chi\rightarrow
\psi \sigma^2  \left(q \,A_j -\tilde g \, B_j \right){\bar \sigma}^j \sigma^2 \psi^\dagger
=\psi^\dagger \left(q \,A_j -\tilde g \, B_j \right){\bar \sigma}^j \psi~,
\eeq
and we see that the interaction terms (\ref{dyonint}) are invariant under CP.

Thus we conclude that to have a CP invariant theory of dyons (in an $SL(2,Z)$ basis where $\theta=0$) the spectrum must contain dyons in pairs with charges $(q,g)$ and $(q,-g)$.
This condition can also be obtained for a theory with bosonic monopoles and/or dyons. However in the case of bosons
due to the absence of chirality it can be equivalently restated as a requirement that every dyon of charge $(q,g)$ is accompanied by a dyon of charge $(-q,g)$.
We note that the requirement of a CP invariant spectrum leads to an interesting consequence.
We can easily see that
given a set of Weyl fermions with charges $(q_j,g_j)$ we find that sums of odd powers of $g_j$ vanish for a CP invariant spectrum, eg.:
\beq
\sum_j g_j&=&0~,\label{CPconstraint1} \\
\sum_j q_j g_j&=&0~,\label{CPconstraint2}  \\
\sum_j g_j^3&=&0~\label{CPconstraint3} .
\eeq
If $\theta \neq 0$, then the pairs should be formed using the effective Witten charge $(q_i+\frac{\theta}{2\pi} g_i,g_i)$. In this case CP invariance would imply the condition
\begin{equation}
\sum_i g_i (q_i+\frac{\theta}{2\pi} g_i)=0\, .
\label{eq:fullCP}
\end{equation}
This condition exactly coincides with the requirement that the $\beta$-function for $\theta$  in Eq.~(\ref{betatheta}) vanishes, and is also $SL(2,Z)$ invariant. If (\ref{eq:fullCP}) does not hold then even if one starts with $\theta =0$ there would be an additive renormalization of $\theta$, implying that CP is an anomalous symmetry.  
Of course if there are any massless charged fermions then $\theta$ is not a  physical parameter, since it can be removed by a chiral rotation of the massless fermion.

\section{The Axial Anomaly}
\label{sec:axial}
\setcounter{equation}{0}
\setcounter{footnote}{0}

\begin{center}
\begin{figure}[htb]
\begin{center}
\includegraphics[width=0.3\hsize]{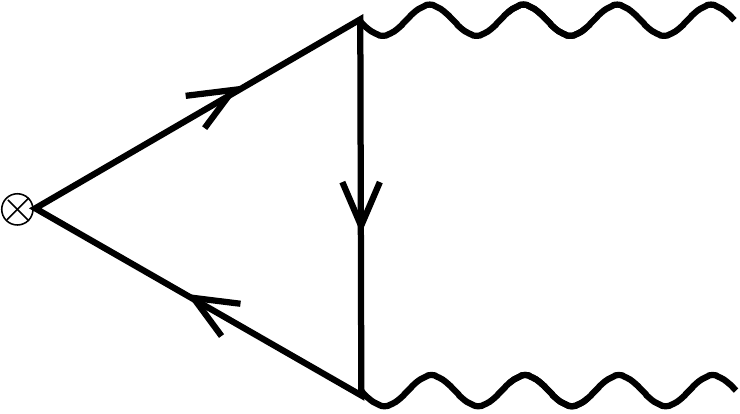}
\end{center}
\caption{The fermion triangle diagram which contributes to the anomaly. One must also
add the crossed graph where the gauge bosons are interchanged.}
\label{Fermiontriangle}
\end{figure}
\end{center}

As a warm-up we will first consider the axial anomaly \cite{ABJ,Fujikawa:1979ay} of a chiral dyon, this can be computed in the Zwanziger formalism \cite{Zwanziger:1970hk} from a triangle diagram with the axial current at one vertex and $U(1)$ gauge fields at the other two vertices.
Since the axial charge of any fermion is just one, we expect in general that the coefficient of the axial anomaly is related to the one-loop $\beta$ function, both of which can be calculated in the Zwanziger formalism (see Fig.~\ref{Fermiontriangle}).

A simpler way of obtaining the anomaly is to follow the method of Argyres and Douglas \cite{ArgyresDouglas} of using $SL(2,Z)$ transformations to map the theory with a dyon to a dual theory with an electric charge,  perform the calculations in the dual theory, and then map back, as we did for the $\beta$-function in Sec.~\ref{sec:SL2Z}.
Thus we want to perform $SL(2,Z)$ transformations of the sort (\ref{eq:tautrans})-(\ref{sl2zcharge}).
As in (\ref{eq:dyontrans}) one can map a dyon with charges $(q,g)$ to a dual electron with charge $n$, where $n$ is the greatest common divisor  of the integers $q$ and $g$, using a transformation with $c=g/n$ and $d=q/n$.
In the dual theory with electric charge $n$, the axial anomaly is
\beq
\partial_\mu j^\mu_A(x) = \frac{n^2}{16 \pi^2} F^{\prime \mu\nu} \,^* F^\prime_{\mu\nu}= \frac{n^2}{32 \pi^2} {\rm Im} \left(  F^{\prime \mu \nu}+ i \,^*F^{\prime \mu \nu}\right)^2~.
\eeq

Using (\ref{fsl2z}) we find that in the original theory with a dyon the axial anomaly is
\beq
\partial_\mu j^\mu_A(x) &=&  \frac{n^2}{32 \pi^2} {\rm Im}\left(c \tau^*+d\right)^2 \left(  F^{\mu \nu}+ i \,^*F^{\mu \nu}\right)^2  \\
&=&
 \frac{1}{16 \pi^2} {\rm Re} \left(q+\tau^* g \right)^2   F^{\mu \nu} \,^*F_{\mu \nu}  +  \frac{1}{16 \pi^2}  {\rm Im}\left(q+\tau^* g \right)^2  F^{\mu \nu} \,F_{\mu \nu}\nonumber\\
 &=& \frac{1}{16 \pi^2}\left\{  \left[\left(q+\frac{\theta}{2 \pi} g\right)^2 \! \!\!-g^2 \frac{16 \pi^2}{e^4}\right]   F^{\mu \nu} \,^*F_{\mu \nu} +   \left[q g+ \frac{\theta}{2 \pi} g^2\right] F^{\mu \nu} \,F_{\mu \nu}\right\}.
 \label{fullaxialanomaly}
\eeq
We immediately recognize that the coefficients are indeed determined by the one-loop $\beta$ function contributions as expected.
The second term, proportional to the gauge kinetic term $F^{\mu \nu} \,F_{\mu \nu}$,  may give one pause. If the theory is CP invariant, then this term is clearly absent. However if it is not CP invariant, one can choose to rotate away $F^2$ in the Lagrangian, instead of $F ^*F$ . This does not mean that we would have a theory without kinetic terms: in the presence of monopoles $F ^*F$ is not a total derivative, and hence it can also serve as a good kinetic term. What one cannot do is rotate $F^2$ and $F ^*F$ away at the same time. If we choose to work in the basis where we have  rotated $\theta$ to zero, we are left with the expression (which is of the form envisioned in~\cite{ArgyresDouglas}):
\beq
\partial_\mu j^\mu_A(x) &=&   \frac{1}{16 \pi^2} \left\{\left[q^2-g^2 \frac{16 \pi^2}{e^4}\right]   F^{\mu \nu} \,^*F_{\mu \nu}  + q g \,F^{\mu \nu} \,F_{\mu \nu}\right\}~.
\eeq
If one wants to ensure that a global $U(1)_X$ symmetry is anomaly free, independently of the renormalization scale (that is ignoring possible scale dependent cancelations between terms with different powers of $e$), then the following three conditions have to be obeyed:
\beq
\sum q_{Xi} q_i^2=0, \ \ \sum q_{Xi} q_i g_i = 0, \ \ \sum q_{Xi} g_i^2 =0.
\eeq
These can be interpreted as separate $U(1)_X U(1)_{el}^2, U(1)_X U(1)_{el} U(1)_{mag}$ and $U(1)_X U(1)_{mag}^2$ anomaly cancelation conditions. The only way to avoid three separate anomaly cancelation conditions is if the gauge coupling is exactly at a fixed point throughout the running, and the charges satisfy 
\beq
\sum q_i^X q_i^2 = \frac{16\pi^2}{e^4} \sum q_i^X g_i^2
\eeq
 for the fixed point coupling $e$. Alternatively it might be possible to have an enhanced global symmetry only at an IR fixed point if the theory runs toward a fixed point and the fixed point coupling satisfies the above equation.

\section{Gauge Anomalies}
\label{sec:mixed}
\setcounter{equation}{0}
\setcounter{footnote}{0}

The most convenient way to phrase the requirement for anomaly cancelation for gauge symmetries is that under an anomalous gauge transformation the Lagrangian will pick gauge dependent terms. For example for a $U(1)$ gauge group (with only electric charge) which has mixed anomalies with an $SU(N)$ gauge group one finds that the following gauge dependent terms appear in the action~\cite{Preskill}
\beq
{\mathcal L}_{\rm anom}&=&c\, \Omega \,G^{a \mu \nu}\,^*G^a_{\mu \nu}
\eeq
where $\Omega$ is the gauge transformation parameter of the the $U(1)$ and $G^{a \mu \nu}$ is the field strength of the $SU(N)$ gauge group.

If we introduce fields magnetically charged under the $U(1)$,  it is again most convenient to use the Zwanziger formalism and introduce the two gauge potentials $A$ and $B$ as we did in section \ref{sec:zwans}. In this case there will be a separate gauge transformation parameter $\Omega_A$ for the $A$-field and $\Omega_B$ for the $B$-field.  The combined gauge parameters
\beq
\Omega= \Omega_A+i \,\Omega_B
\eeq
should transform the same way as the gauge potentials (\ref{gaugepotSL2Z}) under the $SL(2,Z)$ transformation.  Thus, mapping the dyon to an electron and mapping back we obtain an expression for  the gauge varying terms in the Lagrangian:
\beq
{\mathcal L}_{\rm anom}&=&\frac{n\,{\rm Tr}\, T^a(r) T^b(r) }{16 \pi^2}\,  \Omega^\prime_A  \,G^{a \mu \nu}\,^*G^b_{\mu \nu}=\frac{n\,{\rm Tr}\, T^a(r) T^b(r) }{16 \pi^2}\,{\rm Re}\,  \Omega^\prime  \,G^{a \mu \nu}\,^*G^b_{\mu \nu}\\
&=&\frac{n\, T(r) }{16 \pi^2} {\rm Re}\,  \left(c \tau^* +d\right) \Omega \,G^{a \mu \nu}\,^*G^a_{\mu \nu} \\
&=&\frac{T(r)}{16 \pi^2} \, \left[ \left(q+ \frac{\theta}{2 \pi} g \right) \Omega_A+ g \frac{4 \pi}{e^2} \Omega_B\right] \,G^{a \mu \nu}\,^*G^a_{\mu \nu} ~,
\label{mixedanomcalc}
\eeq
where Tr$T^a T^b=T(r) \delta^{ab}$ and $T(r)$ is the Dynkin index. Again $\theta$ can be rotated away,  and the  new anomaly condition, aside from the ordinary mixed anomaly condition (\ref{SUN^2U1anomaly}), corresponding to $SU(N)^2 U(1)_{mag}$ is
\beq
\sum_j T(r_j)  \, g_j =0~.\label{magneticSUN^2anomaly}
\eeq
Similarly  the vanishing of the mixed gravitational anomaly coefficient requires
\beq
\sum_j  \, g_j =0~.\label{magneticgravanomaly}
\eeq

With an additional $U(1)_X$ gauge symmetry in  the theory, there would also be a $U(1)_X U(1)^2$ anomaly as well as a $U(1) U(1)_X^2$  anomaly.  From the preceding examples we see that the anomaly coefficients for $U(1)_X U(1)^2$  can be read off from the axial anomaly, Eq.~(\ref{fullaxialanomaly}):
\beq
\sum_j q_{Xj} \left[q_j^2-g_j^2 \frac{16 \pi^2}{e^4}\right]
\eeq
and
\beq
\sum_j q_{Xj} q_j\, g_j  \label{magneticU1^2anomalyCP}~,
\eeq
where, as before we have again rotated $\theta$ to zero.
Now we see that since $e$ is a running coupling, there are three separate conditions, two of which are new, Eq.~(\ref{magneticU1^2anomalyCP}) and
\beq
\sum_j q_{Xj} \, g_j^2 =0~\label{magneticU1^2anomaly}.
\eeq

The $U(1) U(1)_X^2$  anomaly can be seen as a special case of the calculation in  (\ref{mixedanomcalc}) with $U(1)_X$ charges replacing the $SU(N)$ generators.  Thus we find that the new anomaly cancelation condition in this case is
 \beq
 \sum_j q_{Xj}^2 \, g_j =0~\label{magneticU1anomaly} ~.
 \eeq

 Finally, let us consider the cubic gauge anomaly.
Again mapping the dyon to an electron and mapping back we have:
\beq
{\mathcal L}_{\rm anom}&=&\frac{n^3 }{16 \pi^2}\,  \Omega^\prime_A  \,F^{\prime \mu \nu}\,^*F^\prime_{\mu \nu} = \frac{n^3 }{32 \pi^2}\,{\rm Re}\left[ \Omega^\prime \right] \, {\rm Im}\left[  \left(F^{\prime \mu \nu}+i \,^*F^\prime_{\mu \nu}\right)^2 \right] \\
&=& \frac{n^3 }{32 \pi^2}\, {\rm Re}\left[ \left(c \tau^* +d\right) \Omega \right]  {\rm Im}\left[  \left(c \tau^* +d\right)^2   \,\left(F^{\mu \nu}+i \,^*F_{\mu \nu}\right)^2 \right] \nonumber \\
&=&\frac{1}{16 \pi^2} \,  \left[ \left(q+ \frac{\theta}{2 \pi} g \right)^3 -\left(q+ \frac{\theta}{2 \pi} g \right) \frac{16 \pi^2}{e^4} g^2\right] \Omega_A \,F^{\mu \nu}\,^*F_{\mu \nu} \nonumber \\
&&-\frac{1}{16 \pi^2} \,  \left[ -\left(q+ \frac{\theta}{2 \pi} g \right)^2\frac{4 \pi}{e^2} g +\frac{64 \pi^3}{e^6} g^3\right]  \Omega_B \,F^{\mu \nu}\,^*F_{\mu \nu}   \\
&&-\frac{1}{8 \pi^2} \left[ \left(q+ \frac{\theta}{2 \pi} g \right)^2\frac{4 \pi}{e^2} g\,  \Omega_A
+\left(q+ \frac{\theta}{2 \pi} g \right) \frac{16 \pi^2}{e^4} g^2\,\Omega_B \right] F^{\mu \nu}\,F_{\mu \nu}  ~.
\nonumber
\eeq
Again setting $\theta=0$ we find the new non-trivial anomaly cancelation conditions (corresponding to $U(1)_{el}^2 U(1)_{mag}$, $U(1)_{el}U(1)_{mag}^2$ and $U(1)_{mag}^3$):
  \beq
\sum_j q_j^2  \,g_j &=&0~,\label{newanomalyCP1}\\
\sum_j q_j \,g_j^2 &=&0~,\label{newanomaly}\\
 \sum_j g_j^3 &=&0~.\label{newanomalyCP2}
 \eeq

\section{Conclusions}
We have seen that  in theories with both electric and magnetic charges there are 8 new non-trivial gauge anomaly conditions (even when the $\theta$ parameter vanishes), given in Eqs.~(\ref{magneticSUN^2anomaly}),  (\ref{magneticgravanomaly}), (\ref{magneticU1^2anomalyCP}), (\ref{magneticU1^2anomaly}),  (\ref{magneticU1anomaly}), (\ref{newanomaly}), (\ref{newanomalyCP1}), and (\ref{newanomalyCP2}). We note that  there is an interchange symmetry among the complete set of 13 anomaly conditions (the original five, given in Eqs.~(\ref{cubicanomaly})-(\ref{gravanomaly}) supplemented with the 8 new conditions described above). If we simply interchange $q_j$ and $g_j$, then the set of anomaly conditions is transformed into itself, as we would expect.  This can be simply rephrased as the fact that the full set of anomaly conditions is invariant under S duality where $q_j\rightarrow g_j$ and $g_j\rightarrow -q_j$ . One can also check that the full set is invariant under the T symmetry
$\theta\rightarrow \theta+ 2 \pi$ and $q_j\rightarrow q_j -g_j$. So the set of anomaly conditions is $SL(2,Z)$ invariant.

It would be very interesting to apply these constraints for building new models of electroweak symmetry breaking \cite{wip}.

\section*{Acknowledgements}
We thank   Louis Alvarez-Gaume, Andy Cohen, Michael Dine, Dan Green, Dan Freed, Howie Haber, Jeff Harvey, Markus Luty, Massimo Porrati, Pierre Ramond, and Martin Schmaltz for useful
discussions and comments.  We also thank the Aspen Center for Physics, and JT thanks CERN, where part of this work was completed. 
The research of C.C. has been supported in part by the NSF grant PHY-0757868 and in part by a U.S.-Israeli BSF grant.
Y.S. is supported in part by the NSF grant  PHY-0653656. J.T. is supported by the US Department of Energy grant DE-FG02-91ER40674.

\end{document}